\documentstyle[twoside,titlepage,psfig,12pt]{article}
\begin{document}
\newcommand{\gtsima}{$\; \buildrel > \over \sim \;$}
\newcommand{\ltsima}{$\; \buildrel < \over \sim \;$}
\newcommand{\simgt}{\lower.5ex\hbox{\gtsima}}
\newcommand{\simlt}{\lower.5ex\hbox{\ltsima}}
\newcommand{\hikpc}{{\hbox {$h^{-1}$}{\rm kpc}} }
\newcommand{\himpc}{{\hbox {$h^{-1}$}{\rm Mpc}} }
\newcommand{\0}{ {\scriptscriptstyle {0}} }
\newcommand{\T}{ {\scriptscriptstyle {\rm T}} }
\newcommand{\A}{ {\scriptscriptstyle {\rm A}} }
\newcommand{\kms}{ {\rm km/sec} }
\newcommand{\keV}{ {\rm keV} }
\newcommand{\mpc}{ {\rm Mpc} }
\newcommand{\bfs}{{\mbox{\boldmath $s$}}}
\newcommand{\bfx}{{\mbox{\boldmath $x$}}}
\newcommand{\bfk}{{\mbox{\boldmath $k$}}}
\newcommand{\smbfk}{{\mbox{\footnotesize\boldmath $k$}}}
\newcommand{\smbfx}{{\mbox{\footnotesize\boldmath $x$}}}
\newcommand{\sVert}{{\scriptscriptstyle\Vert}}
\def\pp{\par\parshape 2 0truecm 15.5truecm 1truecm 14.5truecm\noindent}

\begin{titlepage}
\vspace*{-1.5cm}
\begin{minipage}[c]{3cm}
  \psfig{figure=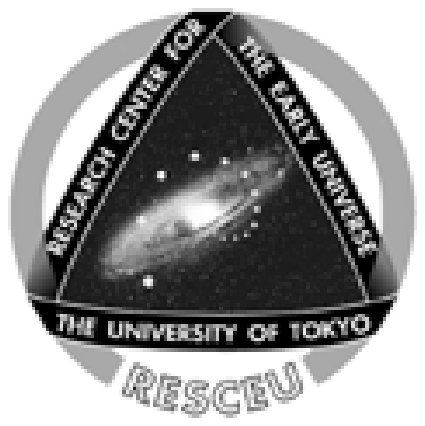,height=3cm}
\end{minipage}
\begin{minipage}[c]{10cm}
\begin{centering}
{
\vskip 0.1in
{\large \sf 
THE UNIVERSITY OF TOKYO\\
\vskip 0.1in
Research Center for the Early Universe}\\
}
\end{centering}
\end{minipage}
\begin{minipage}[c]{3cm}
\vspace{2.5cm}
RESCEU-24/96\\
UTAP-239/96
\end{minipage}\\
\vspace{1cm}

\addtocounter{footnote}{1}
 
\baselineskip=20pt

\begin{center}

  {\Large\bf Cosmological Redshift-Space Distortion II :\\ distance in
    an inhomogeneous universe \\ and evolution of bias}

\vspace{1cm}
 
{\sc  Yasushi Suto and Takahiko Matsubara}

\end{center}
 

\noindent {\it Department of Physics, School
  of Science, The University of Tokyo, Tokyo 113, Japan.}

\noindent {\it Research Center For the Early Universe (RESCEU), 
  School of Science, The University of Tokyo, Tokyo 113, Japan.}


\centerline { e-mail: suto@phys.s.u-tokyo.ac.jp, ~
matsu@phys.s.u-tokyo.ac.jp}

\vspace{7mm}


\vspace{1.0cm}

\baselineskip=14pt

\centerline {\bf Abstract}

  We discuss the two potentially important effects which should be
  taken into account in the analysis of the cosmological
  redshift-space distortion especially at high redshifts; the effect
  of inhomogeneities in the light propagation and the evolution of
  bias.  Although such inhomogeneities affect significantly the degree
  of the cosmological redshift distortion, one can determine the
  density parameter $\Omega_0$ from the distortion pattern at $z=0$,
  and the cosmological constant $\lambda_0$ from that at $z = (1\sim
  2)$. In particular we show that in low-density ($\Omega_0\ll 1$)
  universes one can determine the value of $\lambda_0$ and even the
  evolution of the bias parameter almost insensitively to the degree
  of the clumpiness of the matter distribution in the universe.

\noindent

\medskip

\noindent {\it Subject headings:} cosmology: theory 
--- large-scale structure of the universe --- methods: statistical
\vfill

\end{titlepage}

\section{Introduction}

Alcock \& Paczy\'nski (1979) proposed a geometrical test for the
cosmological constant $\lambda_0$. Only very recently this idea
attracted much attention, and several authors discussed more specific
and realistic methods applicable for samples of galaxies and quasars
at high redshifts $z$ (Phillipps 1994; Ryden 1995; Ballinger, Peacock,
\& Heavens 1996; Matsubara \& Suto 1996, hereafter Paper I).

The angular diameter distance $D_\A$, which plays a key role in the
geometrical test at high $z$, depends sensitively on the inhomogeneous
matter distribution as well as $\lambda_0$ and the density parameter
$\Omega_0$.  A reasonably realistic approximation to the light
propagation in an inhomogeneous universe is given by Dyer \& Roeder
(1973). They assume that the fraction $\alpha$ of the total matter in
the universe is distributed smoothly and the rest is in the clumps. If
the observed beam of light propagates far from any clump, then the
angular diameter distance $D_\A(z;\alpha,\Omega_0,\lambda_0)$
satisfies
\begin{equation}
\label{eq:da}
{d^2 D_\A \over dz^2} 
+ \left[ {2 \over 1+z} + {1\over H(z)} {d H(z) \over dz}\right]
{d D_\A \over dz} 
+ {3 \over 2} {\alpha H_0^2 \Omega_0 (1+z) \over H(z)^2} D_\A = 
0 ,
\end{equation}
with $D_\A(z=0)=0$ and $d D_\A/dz (z=0)= 1/H_0$, where $H_0$ is the
present Hubble constant ($\equiv 100 h$km/sec/Mpc) and $H(z) \equiv
H_0 \sqrt{\Omega_0 (1 + z)^3 + (1-\Omega_0-\lambda_0) (1 + z)^2 +
  \lambda_0}$.  Alcock \& Paczy\'nski (1979) and all the above papers
in the cosmological redshift-space distortion adopted a {\it standard}
distance which corresponds to an extreme case of $\alpha=1$. The
effect of inhomogeneity represented by the parameter $\alpha$ in the
above approximation, however, is quite large for high $z$ (see Figure
\ref{fig:daeta} below).

In addition, the possible time-dependence of the bias parameter
$b(z)$, the ratio of the fluctuations of structures relative to those
of mass, hampers the observational estimate of $\lambda_0$ and
$\Omega_0$ through the anisotropies of the power-spectrum (Ballinger,
Peacock, \& Heavens 1996) and of that of two-point correlation
functions (Paper I). In what follows we adopt the linear bias model by
Fry (1996):
\begin{equation}
   b(z) = 1 + {D(0) \over D(z)} (b_0-1),
\label{eq:fryb}
\end{equation}
where $b_0$ is the present value of the bias parameter, and $D(z)$ is
the linear growth rate. In his model, objects are formed at a fixed
time by a process of biasing and the subsequent motions are determined
by the gravitational potential. The result (\ref{eq:fryb}) is a
consequence of linear pertrubation theory of his model for the
universe with arbitrary $\Omega$ and $\Lambda$.  Apparently this is
one of the simplest possibilities from perturbation consideration, and
more realistically $b(z)$ would be determined by very complicated
astrophysical processes. Nevertheless this model serves to highlight
the effect of the evolution of bias specifically in the present
context.

In this {\it Letter}, we extend the analysis of Paper I by taking
account of the above two effects, and compute the anisotropy of the
two-point correlation function in cold dark matter (CDM) models using
linear theory.

\section{Cosmological redshift distortion in an inhomogeneous universe}

Let us consider a pair of objects located at redshifts $z_1$ and $z_2$
whose redshift difference $\delta z \equiv z_1-z_2$ is much less than
their mean redshift $z \equiv (z_1+z_2)/2$.  Then the observable
separations of the pair parallel and perpendicular to the
line-of-sight direction are given as $s_\sVert = \delta z/H_0$ and
$s_\bot = z\delta\theta/H_0$, respectively, and $\delta\theta$ denotes
the angular separation of the pair on the sky.  The cosmological
redshift-space distortion originates from the anisotropic mapping
between the redshift-space coordinates, $(s_\sVert, s_\bot)$, and the
real comoving ones, $(x_\sVert, x_\bot) \equiv (c_\sVert s_\sVert,
c_\bot s_\bot)$; $c_\sVert (z) = H_0/H(z)$ and $c_\bot(z) = H_0 (1+z)
D_\A/z$.  The cosmological redshift distortion is based on the fact
that $c_\bot(z)$ and $c_\sVert (z)$ depend on $\Omega_0$ and
$\lambda_0$ in a different manner and is characterized by their ratio
$\eta(z) \equiv c_\sVert(z)/c_\bot (z)$.

As shown in Figure \ref{fig:daeta}, the effect of inhomogeneity makes
larger difference than that of $\lambda_0$ especially for $z\gg 1$
(throughout the present paper we compute $D_\A$ using the ANGSIZ
routine provided by Kayser, Helbig, \& Schramm 1996).  Moreover
$\alpha$ is a stochastic variable in principle, and may be different
depending on the line-of-sight direction. Thus $\alpha$ cannot be
predicted theoretically unless one adopts some ad-hoc assumption on
the distribution of dark matter.  In reality, however, the situation
is not so bad. Since the expectation value of $\alpha$ is determined
by the effective volume of the beam of the light bundles, it depends
on the depth $z$ and the angular separation $\delta \theta$ (of the
quasar pair in the present example). For larger $z$ and larger $\delta
\theta$, $ \alpha (z, \delta\theta)$ should approach unity in any
case, and the result based on the standard distance as in Paper I
would be basically correct.  Since we do not have any justifiable
model for $ \alpha (z, \delta\theta)$, we will consider two extreme
cases $\alpha (z, \delta\theta) = 1$ (filled beam) and $0$ (empty
beam). Our main purpose here is to highlight the importance of the
effect even though more realistically $\alpha (z, \delta\theta)$ is
somewhere in between the two extreme cases and $\alpha (z,
\delta\theta)=1$ is a qualitatively reasonable guess especially for
$z\gg1$ and $\delta\theta \gg1$.

It is quite reassuring that even in these extreme cases the
inhomogeneity effect is much weaker than that of $\lambda_0$ at
$z\simlt 2$ in {\it low density} universes as the right panels in
Figure \ref{fig:daeta} illustrate clearly. Since most observational
evidence points to a relatively low value of $\Omega_0$ around
$(0.1\sim 0.3)$ (e.g., Peebles 1993; Suto 1993), this suggests that
the optimal redshift to determine $\lambda_0$ in low $\Omega_0$
universes is $z=(1\sim2)$; the redshift distortion of galaxy
correlation functions at $z\ll 1$ (Kaiser 1987; Hamilton 1992) is a
good probe of $\Omega_0$ insensitive to $\lambda_0$.

\section{Anisotropies in two-point correlation functions}

The relation between the two-point correlation functions of quasars in
redshift space, $\xi^{(s)}(s_\bot, s_\sVert)$, and that of {\it mass}
in real space $\xi^{(r)}(x)$ can be derived in linear theory (Paper I;
Hamilton 1992):
\begin{eqnarray} 
\label{eq:xis}
   &&
   \xi^{(s)}(s_\bot, s_\sVert) = 
   \left(1+{2\over 3}\beta(z) +{1\over5}[\beta(z)]^2\right) \xi_0(x)
   P_0(\mu) 
   -\left({4\over 3}\beta(z)
   +{4\over7}[\beta(z)]^2\right)\xi_2(x) P_2(\mu)
   \nonumber\\
   && \qquad\qquad
   +\frac{8}{35}[\beta(z)]^2 \xi_4(x) P_4(\mu) ,
\end{eqnarray}
where 
\begin{eqnarray} 
\label{eq:xi2l}
\beta(z) \equiv - {1 \over b(z)}\frac{d\ln D(z)}{d\ln (1+z)}, \qquad
\xi_{2l}(x) = {b^2(z) \over 2\pi^2} 
   \int_0^\infty dk k^2 j_{2l}(kx) P(k;z),
\end{eqnarray}
$j_{n}$'s are the spherical Bessel functions, $x \equiv
\sqrt{{c_\sVert}^2 {s_\sVert}^2 + {c_\bot}^2 {s_\bot}^2}$, $\mu\equiv
c_{\sVert} s_{\sVert} /x$, $P_n$'s are the Legendre polynomials.

Thus $\xi^{(s)}(s_\bot, s_\sVert)$ in linear theory crucially depends
on the power spectrum of the {\it mass} fluctuations $P(k;z)$, the
bias parameter for quasars $b(z)$, and $\alpha$ as well as $\Omega_0$
and $\lambda_0$. Clearly it seems very difficult to break the
degeneracy and reliably determine $\lambda_0$, for instance, without
further knowledges or assumptions of the other parameters.  In
principle, one may resort to the extensive curve fits to the observed
correlation functions at different scales $s$ and redshifts $z$ to
estimate the set of parameters. However there is an optimal range of
different $z$ to determine each parameter.

\begin{figure}
\begin{center}
   \leavevmode\psfig{figure=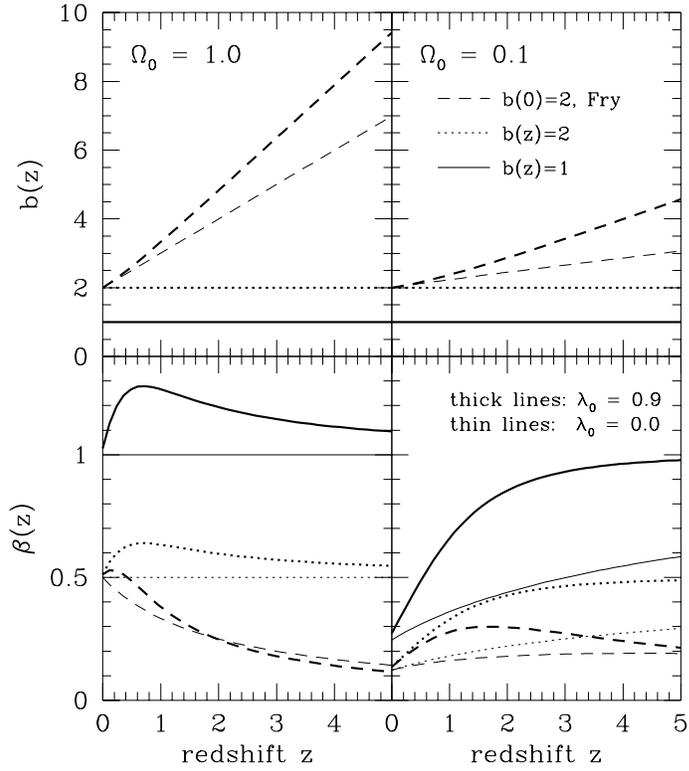,width=12cm}
\end{center}
\caption{
  Evolution of bias and $\beta(z)$. Dashed and solid lines assume
  constant biasing parameter $b(z)=2$ and $1$, respectively, while
  dotted lines adopt the evolution model of Fry (1996) with $b=2$ at
  $z=0$. Thick and thin lines correspond to $\lambda_0=0.9$ and $0$,
  respectively.
\label{fig:bbeta}
}\end{figure}

\begin{figure}
\begin{center}
   \leavevmode\psfig{figure=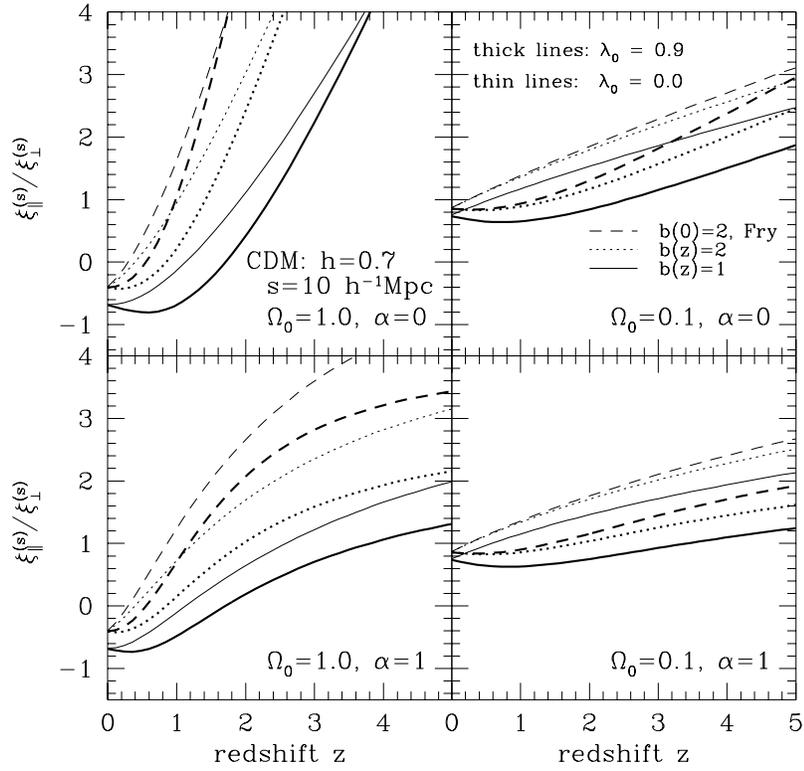,width=12cm}
\end{center}
\caption{The anisotropy parameter
  $\xi^{(s)}_{\sVert}(s)/\xi^{(s)}_\bot(s)$ as a function of $z$ at
  $s=10h^{-1}$Mpc in cold dark matter universes with
  $H_0=70$km/sec/Mpc.  Upper-panels assume $\alpha=0$ (empty beams),
  while lower-panels $\alpha=1$ (filled beams).
\label{fig:anisoz}
}\end{figure}

\begin{figure}
\begin{center}
   \leavevmode\psfig{figure=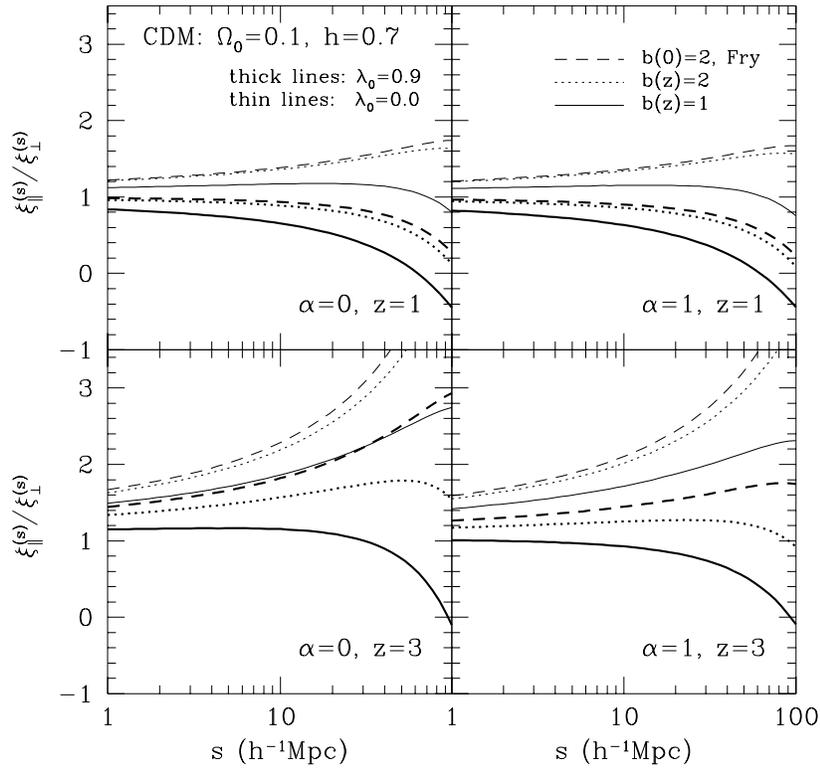,width=12cm}
\end{center}
\caption{The anisotropy parameter
  $\xi^{(s)}_{\sVert}(s)/\xi^{(s)}_\bot(s)$ as a function of $s$ at
  $z=1$ (upper-panels) and at $z=3$ (lower-panels) in cold dark matter
  universes with $H_0=70$km/sec/Mpc.  Left-panels assume $\alpha=0$
  (empty beams), while right-panels $\alpha=1$ (filled beams).
\label{fig:anisos}
}\end{figure}

Figure \ref{fig:bbeta} shows the evolution of bias
(eq.[\ref{eq:fryb}]; upper panels) and of the resulting $\beta(z)$
parameter (lower panels). This implies that as long as Fry's model of
$b(z)$ is adopted, one can distinguish the value of $\lambda_0$
independently of the evolution of bias {\it only in low density
  ($\Omega_0 \ll 1$) models and at intermediate redshifts ($z\simlt
  2$)}. Together with the indication from Figure \ref{fig:daeta} (\S
2), $z=(1\sim2)$ would be an optimal regime to probe $\lambda_0$ at
least in low-density universes.  Figure \ref{fig:anisoz} illustrates
the extent to which this is feasible simply on the basis of the
anisotropy parameter $\xi^{(s)}_{\sVert}(s)/\xi^{(s)}_\bot(s)$ as a
function of $z$, adopting the power spectrum of the CDM models; in
$\Omega_0=1$ models the value of $\alpha$ completely changes the
$z$-dependence of the anisotropy parameter while $\Omega_0=0.1$ models
are fairly insensitive to it. In addition,
$\xi^{(s)}_{\sVert}(s)/\xi^{(s)}_\bot(s)$ for $z\simlt 2$ in
$\Omega_0=0.1$ models is basically determined by the biasing parameter
at $z=0$ and less affected by the evolution of $b(z)$.  Figure
\ref{fig:anisos} shows the scale-dependence of the anisotropy
parameter in $\Omega_0=0.1$ and $h=0.7$ CDM models.  This clearly
indicates that one can distinguish the different $\lambda_0$ and bias
models by analysing the anisotropy of the correlation function at
$z=1$ almost independently of $\alpha$.

\section{Conclusions}

While the importance of the effect of inhomogeneities in the light
propagation (e.g., Dyer \& Roeder 1973; Kayser, Helbig, \& Schramm
1996) is quite well recognized in the study of the gravitational
lensing, for instance, it has not attracted particular attention in
other fields in observational cosmology until very recently. This
would be primarily because this effect becomes important only at
$z\simgt 1$ where previous redshift surveys do not yet provide good
statistical samples for the accurate determination of the cosmological
parameters.  This effect is important also in estimating $H_0$ and
$q_0$, using the SN Ia at $z\simgt 0.5$ (Kantowski, Vaughan \& Branch
1995; Goobar \& Perlmutter 1995) and the Sunyaev - Zel'dovich effect
(Kobayashi, Sasaki \& Suto 1996).

\begin{figure}
\begin{center}
   \leavevmode\psfig{figure=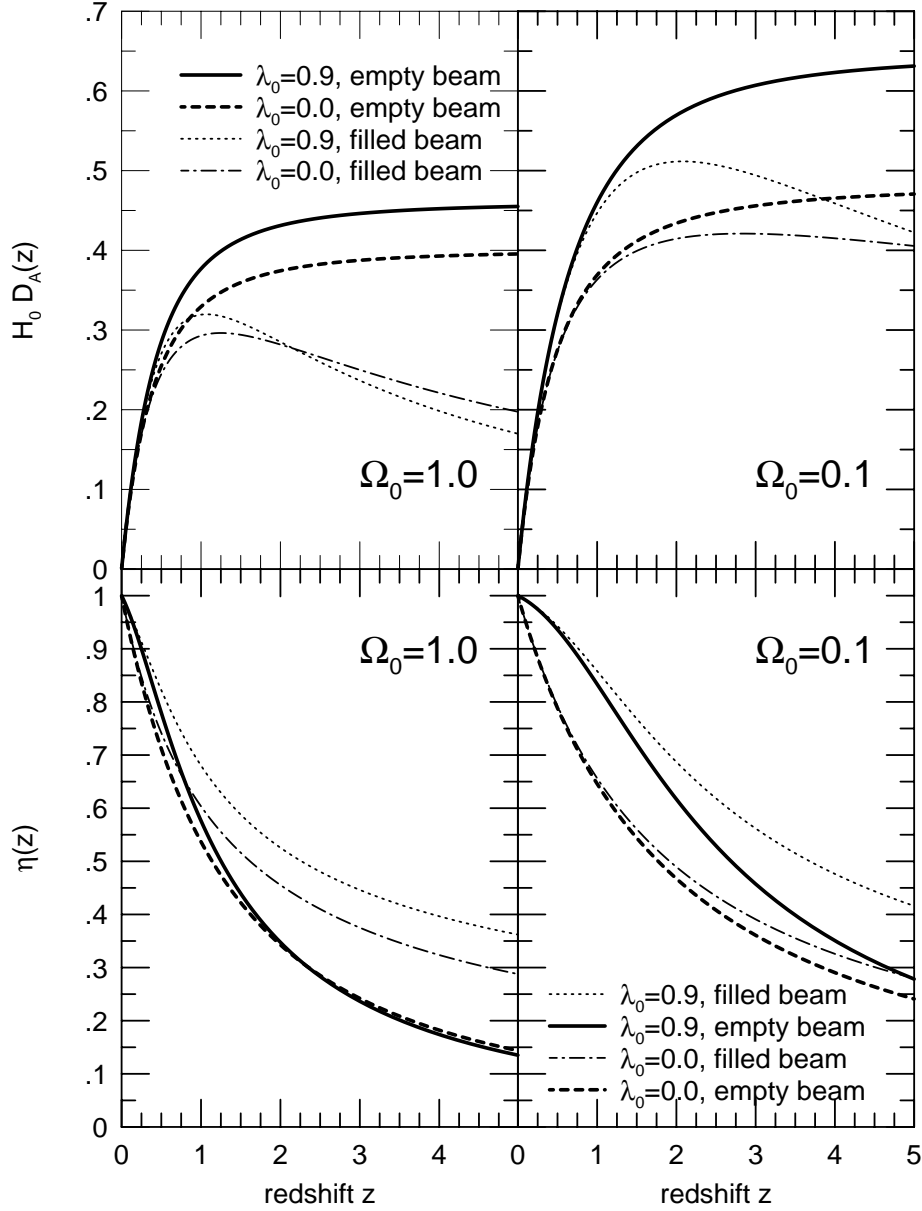,width=12cm}
\end{center}
\caption{
  Effect of inhomogeneity on the angular diameter distance $D_\A(z)$
  (upper panels) and the correction factor $\eta(z)$ (lower-panels)
  for $\lambda_0=0$ and $0.9$ models in $\Omega_0=1$ (left panels) and
  $\Omega_0=0.1$ (right panels) universes. Thick lines indicate the
  results for the empty beam ($\alpha=0)$, while thin lines for the
  filled beam ($\alpha=1)$.
\label{fig:daeta}
}\end{figure}

We have shown that such inhomogeneities, which were neglected in the
original proposal by Alcock \& Paczy\'nski (1979), also affect
significantly the degree of the cosmological redshift distortion, and
in fact hampers the accurate estimate of the cosmological constant
especially in $\Omega_0=1$ universes. In low-density universes,
however, one can determine $\lambda_0$ in principle by optimizing the
range of $z$ (see Fig.\ref{fig:daeta}); $\Omega_0$ (or $\beta$
parameter) is best determined from the redshift-space clustering at
$z=0$ (Kaiser 1987; Hamilton 1992). Once $\Omega_0$ is fixed, the
cosmological redshift distortion around $z\sim 1$ is most sensitive to
the value of $\lambda_0$. Then the distortion at the higher redshift
provides clue to the degree of the clumpiness or $\alpha$ which in
fact would be also a function of $z$.  Since either $\alpha=0$ or
$\alpha=1$ is an extreme case, the realistic situation would be more
favorable to the determination of $\Omega_0$ and
$\lambda_0$. Nevertheless the degree of the clumpiness is another
important factor which should be always kept in mind in the analysis
of the clustering of structures at high redshifts.

\bigskip

We thank Takahiro T. Nakamura, Shin Sasaki, Edwin Turner and David
Weinberg for discussions.  The computation of $D_\A$ is done with the
ANGSIZ routine which was made publicly available by R. Kayser,
P. Helbig, \& T. Schramm.  This research was supported in part by the
Grants-in-Aid by the Ministry of Education, Science, Sports and
Culture of Japan (07CE2002) to RESCEU (Research Center for the Early
Universe), the University of Tokyo.

\bigskip 
\bigskip
\newpage

\parskip2pt
\bigskip
\centerline{\bf REFERENCES}
\bigskip

\def\apjpap#1;#2;#3;#4; {\pp#1, {#2}, {#3}, #4}
\def\apjbook#1;#2;#3;#4; {\pp#1, {#2} (#3: #4)}
\def\apjppt#1;#2; {\pp#1, #2.}
\def\apjproc#1;#2;#3;#4;#5;#6; {\pp#1, {#2} #3, (#4: #5), #6}

\apjpap Alcock, C. \& Paczy\'nski, B. 1979;Nature;281;358;
\apjppt Ballinger, W.E., Peacock, J.A. \& Heavens, A.F. 1996;MNRAS, in
press;
\apjpap Dyer, C.C. \& Roeder, R.C. 1973;ApJL;180;L31;
\apjpap Fry, J. 1996;ApJ;461;L65;
\apjpap Goobar, A. \& Perlmutter, S. 1995;ApJ;450;14;
\apjpap Hamilton, A.J.S. 1992;ApJL;385;L5;
\apjpap Kaiser, N. 1987;MNRAS;227;1;
\apjpap Kantowski, R., Vaughan, T., \& Branch, D. 1995;ApJ;447;35;
\apjppt Kayser, R., Helbig, P., \& Schramm, T. 1996;A\&A,in press;
\apjppt Kobayashi, S., Sasaki, S., \& Suto, Y. 1996;ApJ, submitted;
\apjppt Matsubara,~T. \& Suto, Y.  1996;ApJL, in press (Paper I);
\apjbook Peebles,P.J.E. 1993; Principles of Physical Cosmology;
    Princeton University Press;Princeton;
\apjpap Phillipps, S. 1994;MNRAS;269;1077;
\apjpap Ryden,B.1995;ApJ;452;25;
\apjpap Suto, Y.  1993;Prog.Theor.Phys.;90;1173;

\bigskip
\bigskip
\newpage

\end{document}